\documentclass[a4paper]{aa}
\usepackage{epsfig}
\usepackage{psfig}
\def\ergsec{\hbox{erg s$^{-1}$}}

\def\degmark{^\circ}

\def \rsun {\ifmmode$R$_{\odot}\else R$_{\odot}$\fi}
\def \nh {N${\rm _H}$}
\def \hcm {\hbox {\ifmmode $ atoms cm$^{-2}\else atoms cm$^{-2}$\fi}}
\def \src {1E\thinspace2259+586}
\def \snr {G109.1$-$1.0}
\def\approxgt{\mathrel{\hbox{\rlap{\lower.55ex \hbox {$\sim$}}
        \kern-.3em \raise.4ex \hbox{$>$}}}}
\def\approxlt{\mathrel{\hbox{\rlap{\lower.55ex \hbox {$\sim$}}
        \kern-.3em \raise.4ex \hbox{$<$}}}}
\newcommand {\rosat} {{\it ROSAT}}
\newcommand {\einstein} {{\it Einstein}}

\newcommand {\asca} {{\it ASCA}}

\newcommand {\sax} {{\it BeppoSAX}}

\newcommand {\Msun} {{{\rm M$_{\odot}$}}}
\newcommand {\degree} {$^{\circ}$}
\def\arcmin{\hbox{$^\prime$}}
\def\arcsec{\hbox{$^{\prime\prime}$}}

\newcommand {\chisq} {$\chi ^{2}$}
\newcommand {\rchisq} {$\chi_{\nu} ^{2}$}

\begin{document}

\thesaurus{ (08.14.1; 08.16.7; 09.09.1; 13.25.4; 13.25.5)}

\title{A \sax\ observation of the X-ray pulsar \src\ and the supernova
remnant \snr\ (CTB\thinspace109)}

\author{A.N. Parmar\inst{1} \and T.~Oosterbroek\inst{1} \and F. Favata\inst{1}
\and S. Pightling\inst{2} \and M.J. Coe\inst{2} \and S. Mereghetti\inst{3}
\and G.L. Israel\inst{4}}

\institute
{Astrophysics Division, Space Science Department of ESA, 
ESTEC, P.O. Box 299, 2200 AG Noordwijk, The Netherlands
\and
Department of Physics \& Astronomy, The University, Southampton SO17 1BJ, UK
\and
IFCTR, via Bassini 15, I-20133 Milano, Italy
\and
Osservatorio Astronomico di Roma, via dell'Osservatorio 1, I-00040, 
Monteporzio Catone, Italy}
\offprints{A.N. Parmar: aparmar@astro.estec.esa.nl}
\date{Received ; accepted}
\maketitle

\markboth{\src\ and \snr}{\src\ and \snr}

\begin{abstract}
  
  The 7~s X-ray pulsar \src\ and the supernova remnant (SNR) \snr\ 
  (CTB\thinspace109) were observed by \sax\ in 1996 November.  The
  pulse period of $6.978914 \pm 0.000006$~s implies that \src\ 
  continues its near constant spin-down trend.  The 0.5--10~keV pulse
  shape is characterized by a double peaked profile, with the
  amplitude of the second peak $\sim$50\% of that of the main peak.
  The pulse profile does not exhibit any strong energy dependence.  We
  confirm the \asca\ discovery of an additional low-energy spectral
  component from \src. This can best be modeled as a 0.44~keV
  blackbody, but we cannot exclude that some, or all, of this emission
  arises from the part of the SNR that lies within the pulsar's
  extraction region.
  
  The spectrum of \snr\ is well fit with a non-equi\-librium
  ionization plasma model with a best-fit temperature of $0.95$\,keV.
  The derived mass for the X-ray emitting plasma ($\sim$15--20\,\Msun)
  and its near cosmic abundances imply that the X-ray emission comes
  mainly from mildly enriched, swept-up circumstellar material.  The
  spectrum is strongly out of equilibrium with an ionization age of
  only 3000~yr. This age is in good agreement with that derived from
  hydrodynamic simulations of the SNR using the above X-ray
  temperature.

\end{abstract}


\section{Introduction}
\label{sec:introduction}

\src\ is an unusual pulsar with a pulsation period of 7~s and a very
steep spectrum (photon index, $\alpha$, $\sim$4). The source has
exhibited a near constant spin-down trend at $\sim$5$\times
10^{-13}$ s\,s$^{-1}$ since its discovery. \src\ is located in, or
near, the supernova remnant (SNR) \snr\ which is also known as
CTB\thinspace109. The SNR consists of an approximately hemispherical
shell of X-ray and radio emission and a jet-like lobe located mid-way
between the shell and the pulsar (Gregory \& Fahlman 1980; Hughes et
al. 1981; Morini et al. 1988; Hurford \& Fesen 1995).  There may be a
compact (2--3$'$ radius) synchrotron nebula surrounding the pulsar
(Rho \& Petre 1997).  The location and shape of the X-ray lobe led to
speculation that it is a ``jet'' of material connecting the pulsar and
the SNR (Gregory \& Fahlman 1983). However, its thermal spectrum and
detailed morphology do not support this view (Hurford \& Fesen 1995;
Rho \& Petre 1997).

\src\ together with other sources such as 1E\thinspace1048.1$-$593
and 4U\thinspace0142+614 belong to a small group of pulsars with
similar spin periods (around 6~s) and properties that clearly 
distinguish them from the ``classical'' pulsars in high-mass X-ray
binaries (Mereghetti \& Stella 1995).
Koyama et al. (1987) argue that the slow spin-down
rate implies that \src\ contains an accreting neutron star rotating at 
close to its equilibrium period with a magnetic field of
$\sim$$5\times 10^{11}$~G.
However, there is no direct evidence that the pulsar is
in a binary system. In particular, no optical 
or radio counterparts have been found (Coe \& Jones 1992; Coe et al. 1994) 
and no Doppler shifts in pulse period have been detected 
(Koyama et al. 1989; Hanson et al. 1988; Morini et al. 1988). 

The lack of an optical counterpart and orbital Doppler shifts argue
against a binary models for \src, unless the companion has an
extremely low mass.  Models based on single stars have been proposed,
such as spin-down of a white dwarf (Paczynski 1990; Usov 1994), loss
of magnetic energy of a strongly magnetized neutron star (Thompson \&
Duncan 1993) and a neutron star accreting from a circumstellar disk
(Corbet et al. 1995; van Paradijs et al. 1995).  The detection by
Mereghetti (1995) of a large change in the spin-down rate of the
similar system 1E\thinspace1048.1$-$5937, as well as the long term
pulse frequency fluctuations in \src\ (Baykal \& Swank 1996) support
an accretion hypothesis.
 
We present a detailed study of the X-ray emission from \src\ and \snr\ 
using the imaging capabilities of \sax\ to separate the contributions
of the individual sources.  This topic is particularly interesting
because of the disparate spectral results reported for \src\ as well
as of the uncertain origin of the X-ray lobe in \snr.

\begin{figure*}
\centerline{\psfig{figure=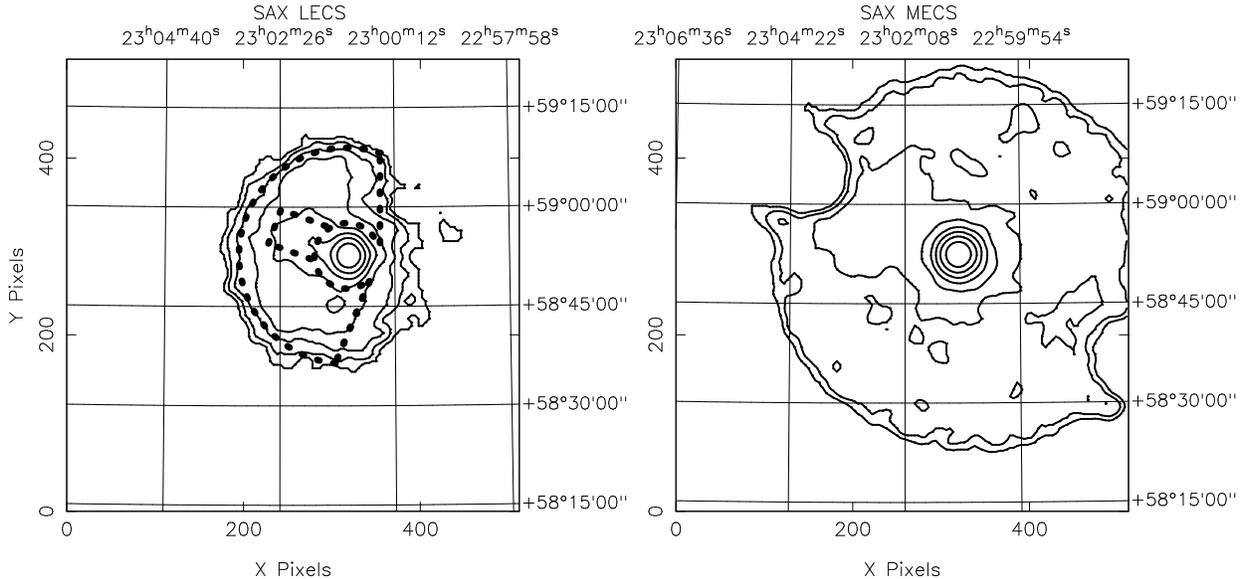,height=18.cm,angle=-90}}
\caption{LECS (left panel) 0.5--2.0~keV and MECS (right panel)
  2.0--8.0 keV
smoothed images of the region containing \src\ and \snr. 
The contours levels represent
1, 2, 3, 6, 12, 25 and 50\% of the peak flux of each image. 
The ``cut-outs'' at the upper-left and lower-right of the MECS image are
due to the removal of instrument calibration source events.
The softness of the SNR and lobe can be seen by the greater asymmetry
of the LECS contours, compared to the MECS. The extraction region used
for the SNR shell spectrum is indicated by the heavy dotted line, 
minus the box shaped region
(which is the extraction region for the lobe spectrum)}
\label{fig:contours}
\end{figure*}

\section{Observations}
\label{sec:observations}

Results from the the Low-Energy Concentrator Spectrometer (LECS;
0.1--10~keV; Parmar et al. 1997) and Medium-Energy Concentrator
Spectrometer (MECS; 1.3--10~keV; Boella et al. 1997) on-board \sax\ 
are presented.  The MECS consists of three identical grazing incidence
telescopes with imaging gas scintillation proportional counters in
their focal planes.  The LECS uses an identical concentrator system as
the MECS, but utilizes an ultra-thin (1.25~$\mu$m) entrance window and
a driftless configuration to extend the low-energy response to
0.1~keV. The fields of view (FOV) of the LECS and MECS are circular
with diameters of 37\arcmin and 56\arcmin, respectively.  In the
overlapping energy range, the position resolution of both instruments
is similar and corresponds to 90\% encircled energy within a radius of
2\farcm5 at 1.5~keV. At lower energies, the encircled energy is
proportional to ${\rm E^{-0.5}}$.  The LECS 0.1-10~keV background
counting rate is $9.7 \times 10^{-5}$~arcmin$^{-2}$~s$^{-1}$.
 
The region of sky containing \src\ and \snr\ was observed by \sax\
between 1996 November 16 03:32 and November 17 05:19~UTC. 
Good data were selected from intervals when the minimum elevation angle
above the Earth's limb was $>$$4\degmark$ and when the instrument
configurations were nominal using the SAXDAS 1.1.0 data analysis package.  
This gives exposures of 53.4~ks for the MECS and 15.2~ks
for the LECS, which was only operated during satellite night-time.
The LECS and MECS images, smoothed using a Gaussian filter of width 1\farcm5, 
are shown in Fig.~\ref{fig:contours}. \src\ is visible as a point source close
to the center of both images.
The position of \src\ derived by summing the LECS and the MECS data is 
23$^{\rm h}~01^{\rm m}~06\fs8$, 
58\degree~52\arcmin~16\arcsec\ (J2000.0) 
with a 68\% confidence uncertainty radius of 30\arcsec.
This is consistent with the \einstein\ High Resolution Imager
position of Fahlman et al. (1982).
The SNR is visible as extended
emission throughout both images. 

Separate spectra were extracted for the pulsar, SNR shell and X-ray lobe 
regions. All spectra were rebinned to have $>$20 counts in each bin to allow
the use of $\chi^2$ statistics.  Background subtraction was performed
using standard blank field exposures.

\section{Results}
\subsection {The \src\ spectrum}
\label{subsec:src_spectrum}

LECS and MECS spectra were obtained centered on the pulsar position
using an extraction radius of 4\arcmin.  The \src\ count rates above
background are 0.28 and 0.55~s$^{-1}$ in the LECS and MECS,
respectively.  Examination of the LECS spectrum reveals that the
pulsar is only detected between 0.5 and 8.0~keV and data outside this
range are excluded. Similarly, the MECS fit is restricted to the
energy range 1.65--10~keV.

\begin{figure}
\centerline{\psfig{figure=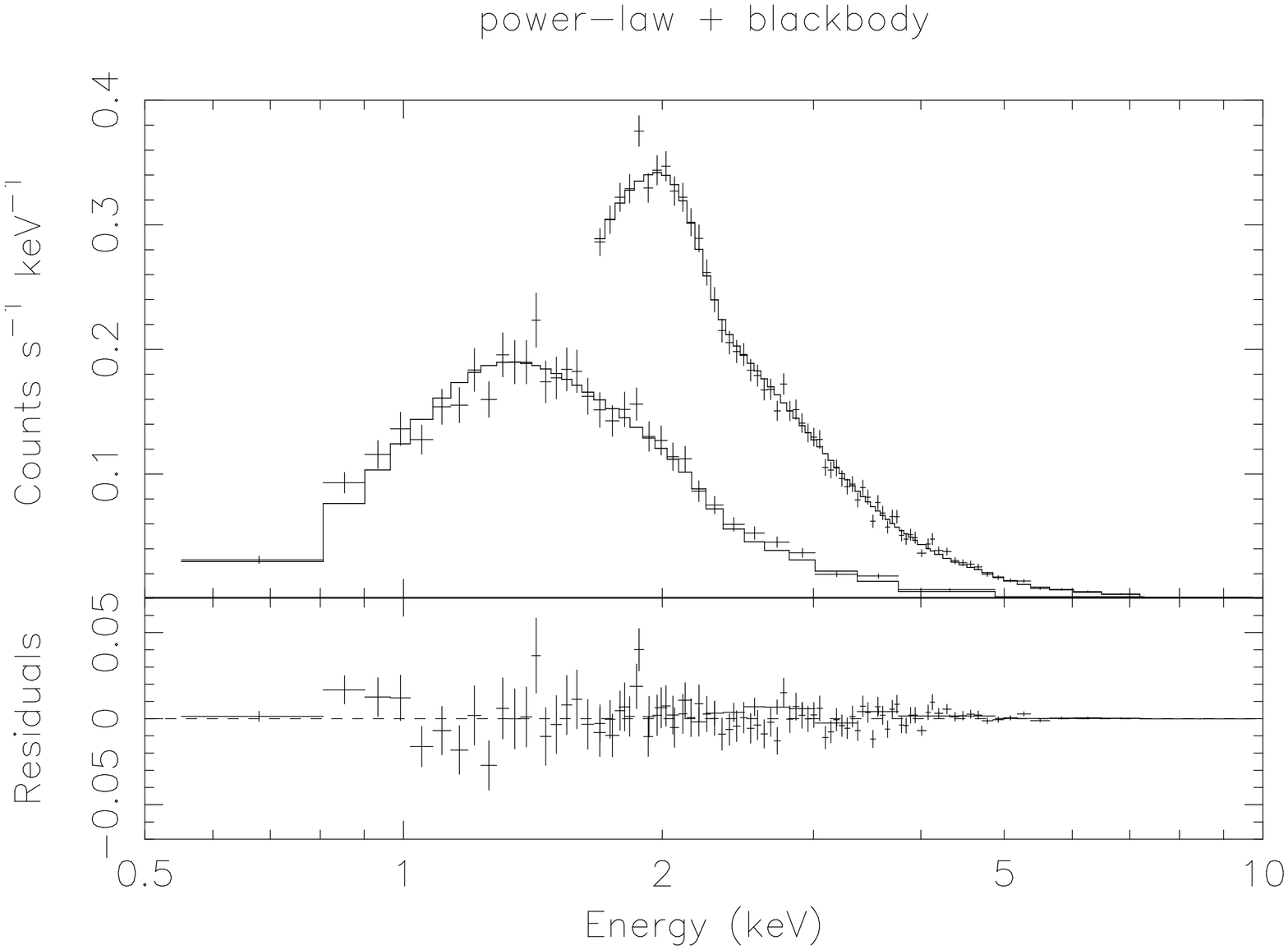,width=6.8cm,angle=-90}}
\centerline{\psfig{figure=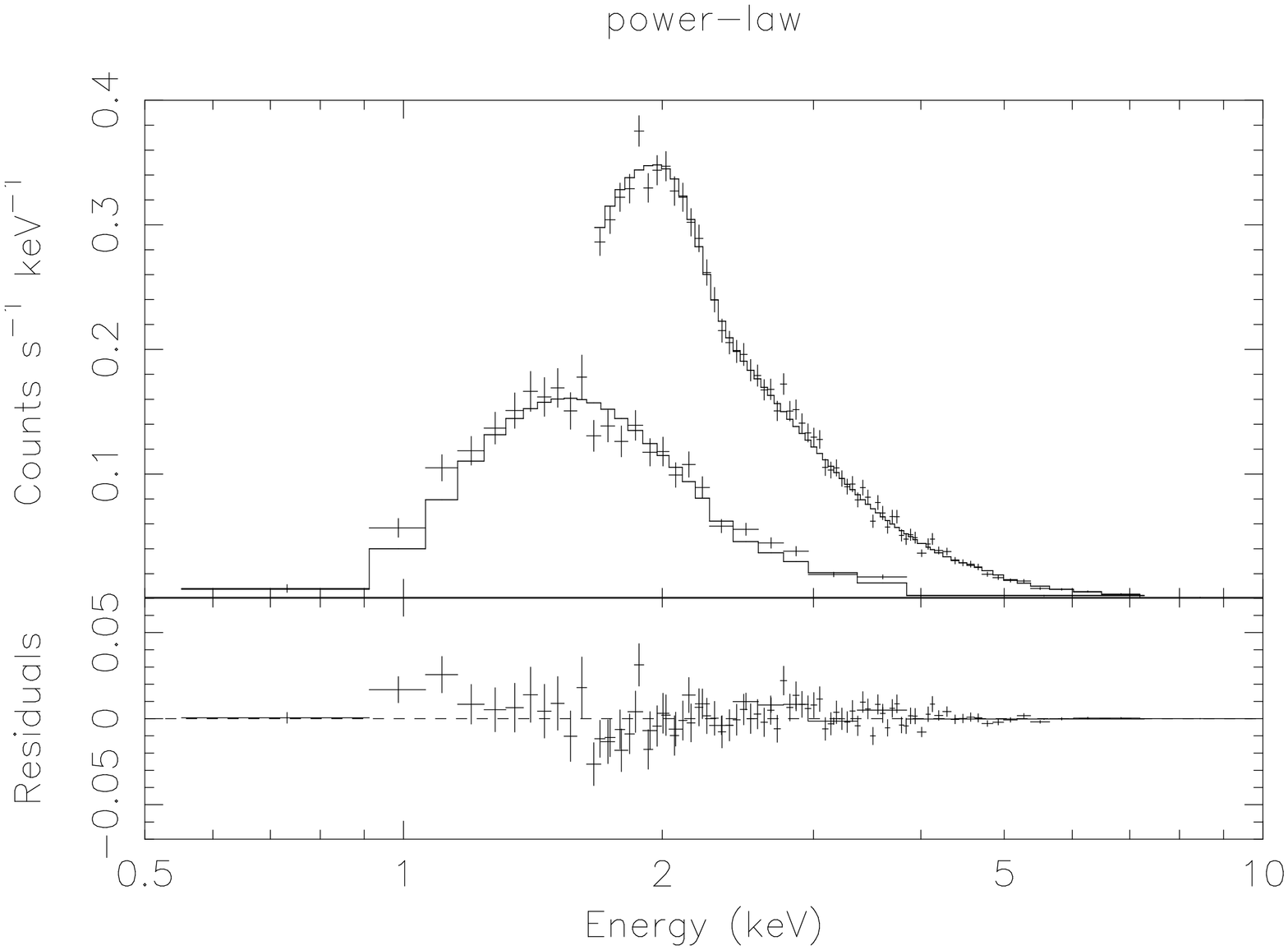,width=6.8cm,angle=-90}}
\caption[]{The best-fit power-law and blackbody fit to the pulsar spectrum 
(top) and the best-fit power-law model fitted to the
pulsar spectrum with 0.19 of the SNR shell spectrum subtracted (bottom).
The smaller panels give the residuals in counts~s$^{-1}$~keV$^{-1}$. 
The models are described in the text and summarized
in Table~\ref{tab:pulsar_fits}}
\label{fig:pulsar_spec}
\end{figure}

The combined LECS and MECS spectrum of the pulsar was first fit 
using a power-law model with 
$\alpha = 4.49 \pm 0.03$ and low-energy absorption 
of $(1.46 \pm 0.05) \times 10^{22}$~\hcm\ yielding
a \chisq\ of 449 for 266 degrees of freedom (dof). 
The photo-electric absorption coefficients of Morisson \& McCammon (1983)
and the solar abundances of Anders \& Grevesse (1989) were used
for all fits.
Examination of the residuals reveals significant 
structure below $\sim$1.0~keV.
A better fit (\chisq\ of 271 for 264 dof)
is obtained when a blackbody component is added.
The spectral parameters
listed in Table~\ref{tab:pulsar_fits} 
are similar to those obtained using the same spectral model
with \asca\ spectra by Corbet et al. (1995) and with a combination of
\asca, \rosat, and {\it BBXRT} spectra by Rho \& Petre (1997).
The remaining residuals near 1~keV may indicate an even more complex
spectral shape.

\begin{figure}
\centerline{\psfig{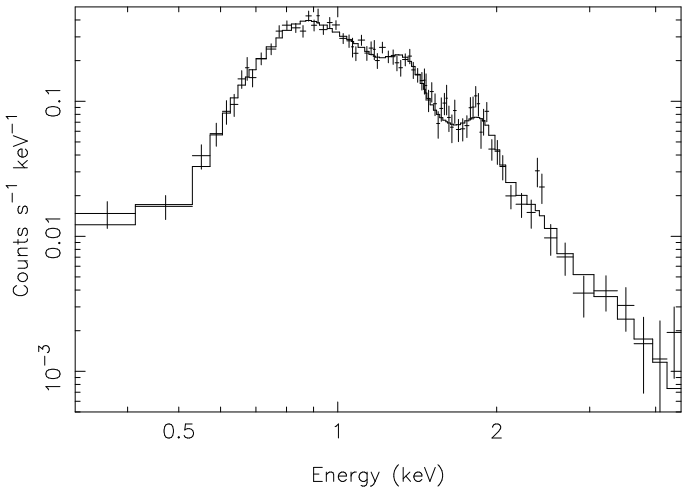}} 
\caption[]{The \snr\ shell spectrum, together with the best-fit
single-component NEI model}
\label{fig:snr_spec}
\end{figure}

In order to investigate an alternative explanation for the low energy
residuals, the spectrum of the SNR shell multiplied by a scale factor, F, was 
subtracted from the \src\ LECS spectrum (the SNR hardly contributes
in the MECS energy range, see Fig.~\ref{fig:snr_spec}).  
The fit with a single power-law  was then repeated and the best-fit value of F 
of $0.19\pm 0.01$ 
determined by minimizing \chisq. This gives a $\chi^2$ of 296 for
265 dof, and the amplitude of the residuals $\approxlt$1~keV is 
similarly reduced as with the power-law and blackbody fit 
(Fig.~\ref{fig:pulsar_spec}). 
The best-fit value of $\alpha$ is $4.86 \pm 0.04$ and \nh\ 
is $(2.18 \pm 0.07) \times 10^{22}$~\hcm. 
The ratio of the areas of the pulsar and SNR shell extraction regions 
(39 and 361~arcmin$^2$, respectively) is 0.11. This is comparable with 
the value obtained for F of 0.19, given the observed variations in 
surface brightness of the SNR, the presence of the X-ray lobe, 
and the extended (2--3$'$ radius) emission around \src\ reported
in Rho \& Petre (1997). 
Although the fit quality is not as good as with the power-law and 
blackbody model, we cannot exclude the possibility that some, or all, 
of the residuals below $\sim$1~keV are caused by the contribution of
the SNR that lies within the pulsar's extraction region.
The \src\ spectral fit results are 
summarized in Table~\ref{tab:pulsar_fits}. The 2--10~keV luminosity
of \src\ is $3.3 \times 10^{34}$~\ergsec\ for a distance of 4~kpc,
identical to the value in Corbet et al. (1995).

\begin{table}
\caption[]{Spectral fit results for \src. 
Uncertainties are given at 68\% confidence}
\begin{flushleft}
\begin{tabular}{ll}
\hline\noalign{\smallskip}
Parameter & Value \\
\noalign {\smallskip}
\hline\noalign {\smallskip}
Pulsar (power-law model)         \\
$\alpha$              & $4.49 \pm 0.03$\\
\nh\ ($10^{22}$~\hcm) & $1.46 \pm 0.05$ \\
\chisq                & 449/266 \\
\noalign{\smallskip}
\hline
\noalign{\smallskip}
Pulsar (power-law \& blackbody model)         \\
$\alpha$              & $3.93 \pm0.09$\\
kT${\rm _{bb}}$ (keV) & $0.44\pm0.01$ \\
Blackbody radius$^a$ (km) & $3.3\pm^{0.2}_{0.3}$ \\
\nh\ ($10^{21}$~\hcm) & $8.7 \pm 0.5$ \\
\chisq                & 271/264 \\
\noalign{\smallskip}
\hline
\noalign{\smallskip}
Pulsar-0.19.SNR (power-law model)\\
$\alpha$              & $4.86 \pm 0.04$ \\ 
\nh\ ($10^{22}$~\hcm) & $2.18 \pm 0.07$ \\
\chisq                & 296/265 \\
\noalign{\smallskip}
\hline
\noalign{\smallskip}
\multicolumn{2}{l}{\footnotesize $^a$For a distance of 4~kpc}
\end{tabular}
\end{flushleft}
\label{tab:pulsar_fits}
\end{table}

\subsection {\src\ pulse timing and phase resolved spectroscopy}
\label{subsec:src_pulsetiming}

\begin{figure}
\centerline{\psfig{figure=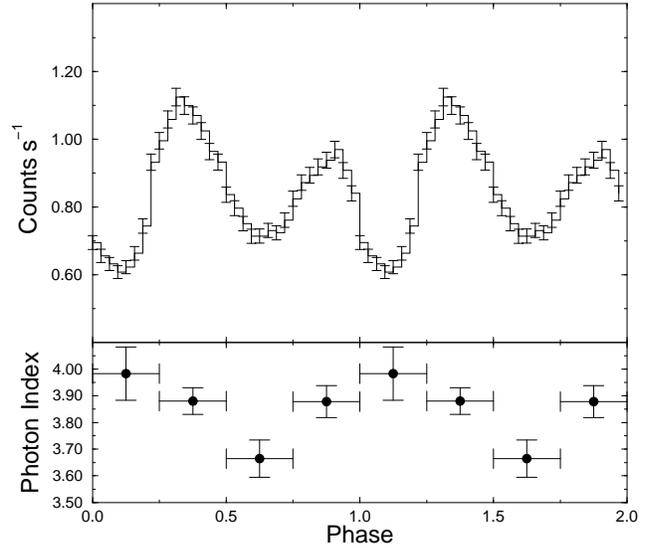,width=8cm,angle=-90}}
\caption[]{The pulse profile (top panel) of \src\ in the 0.5--10~keV
energy range. The bottom panel shows the values of $\alpha$.
The values are repeated for clarity} 
\label{fig:profile}
\end{figure}

The MECS counts were used to determine the \src\ 
pulse period, after correction of their 
arrival times to the solar system barycenter. 
The data were divided
into 14 time intervals (each with $\sim$2000 counts) and for 
each interval the relative phase of the
pulsations determined. This was performed by folding the counts    
at half the pulse period value, in order to obtain light curves
with a stronger modulation. The phases of the 14 time intervals
were then fitted with a linear function giving a best-fit period
of $6.978914 \pm 0.000006$~s. 
The \src\ light curve  (Fig.~\ref{fig:profile})
shows a double-peaked profile with the amplitude of the second
peak about half that of the main peak.

A set of four phase-resolved spectra of the pulsar were
accumulated, approximately coinciding with the peaks and valleys of the
pulse profile. These 
spectra were fit with the same power-law plus blackbody model as
used in Sect.~\ref{subsec:src_spectrum}, with \nh\ fixed at the 
phase-averaged best-fit value.
There are insufficient counts to simultaneously constrain both the 
power-law and blackbody components.
Initially, the blackbody spectral parameters were fixed at their 
phase-averaged best-fit values and only the power-law parameters
allowed to vary. The fits were then repeated with the power-law
parameters fixed while the blackbody parameters were allowed to vary. 
With the latter approach, the two fits in the valleys are unacceptable 
with \rchisq's of 2.9 and 2.3 for 117 dof.
This is because the power-law component contributes
too much flux, even if the contribution from the
blackbody is set to zero. 

The best-fit values of $\alpha$ obtained with the first approach 
are shown in Fig.\ \ref{fig:profile} and reveal a small
phase dependence.
This variation corresponds to a \chisq\/ of 9.4 for 3 dof 
with respect to a constant value. Although this is significant 
at $>$99\% confidence, we cannot be certain that there are
real variations in $\alpha$, since there is no clear
correlation with the flux (as might be expected). 
In addition, small uncertainties in background 
subtraction, or a contribution from the SNR could cause such an effect, 
and so we prefer to set a 
limit of $\pm 0.2$ to any phase-dependent change in $\alpha$.

\subsection{The \snr\ spectrum}
\label{subsec:snr_spectrum}

In order to accumulate the spectrum of the SNR shell a complex shaped
extraction region consisting of a circle, delimited by two straight
lines with the pulsar and jet-like X-ray lobe cut out was used (see
Fig.~\ref{fig:contours}).  Since the spectra of both the shell and the
lobe are soft (there is little flux $\approxgt$2~keV), only
LECS data were used.  The effective area of the LECS depends on
source position within the FOV and the appropriate response matrix was
determined by using 11 point sources whose position and relative
intensities were chosen to mimic the observed count distribution.
Examination of the extracted spectrum shows that the SNR is only
detected between 0.25--4.5~keV and data outside this range are
excluded.

The spectrum of the SNR shell was fit with the Non-Equi\-li\-brium
Ionization (NEI) plasma emission model (plus absorption) implemented
in V.~1.10 of the SPEX package.  Fits were performed with freely
varying O, Ne, Mg, Si, S, Fe and Ni abundances.  A one-component NEI
spectrum satisfactorily describes the spectrum with a \chisq\ of 65
for 66 dof (see Table~\ref{tab:snr_fits}).  The spectrum of the X-ray
lobe was separately extracted and analyzed, using the region indicated
in Fig.~\ref{fig:contours}.  Since the size of the extraction region
is comparable to the standard LECS extraction region, no correction
for source extent was applied to the response matrix.  A background
spectrum was obtained from the standard blank field exposures using
the same extraction region as for the source spectrum.  When the X-ray
lobe spectrum is fit with the same NEI model, the best-fit values are
in all cases within $1\sigma$ of those of the shell spectrum. This is
in agreement with the results of Rho \& Petre (1997) and supports the
view that the two regions are physically related.

\begin{table}[htbp]
\caption[]{Spectral parameters for the one-component NEI fit to the
  SNR shell spectrum. Uncertainties are given at 90\% confidence
  ($\Delta \chi^2 = 2.71$). A distance of 4~kpc is
  assumed.  The X-ray lobe has consistent spectral parameters, except for an
  emission measure (${\rm n_e\,n_H\, V}$) of $27 \times
  10^{56}$\,cm$^{-3}$}
\begin{flushleft}
\begin{tabular}{lll}
\hline\noalign{\smallskip}
Parameter & Best-fit& 90\% conf. \\
          & value & range \\
\noalign {\smallskip}
\hline\noalign {\smallskip}
\nh\ ($10^{21}$~\hcm)    & 6.9 & 5.7--7.6 \\[1pt]
kT (keV)             & 0.95  & 0.68--1.60\\
${\rm n_e\,n_H\, V}$ ($10^{56}$\,cm$^{-3}$) & 63 & 27--120 \\
${\rm n_e\,t}$ ($10^3$\,cm$^{-3}$\,yr) & 1.2  & 0.7--2.4 \\[1pt]
\mbox{[O/H]}  & 2.0 & 0.5--5.0 \\
\mbox{[Ne/H]} & 1.1 & 0.0--1.2 \\
\mbox{[Mg/H]} & 2.4 & 1.3--3.9 \\
\mbox{[Si/H]} & 1.1 & 0.7--1.8 \\
\mbox{[S/H]}  & 0.8 & 0.1--2.0 \\
\mbox{[Fe/H]} & 1.9 & 1.3--2.9 \\
\noalign{\smallskip}
\hline
\noalign{\smallskip}
\end{tabular}
\end{flushleft}
\label{tab:snr_fits}
\end{table}

\begin{figure}
\centerline{\psfig{figure=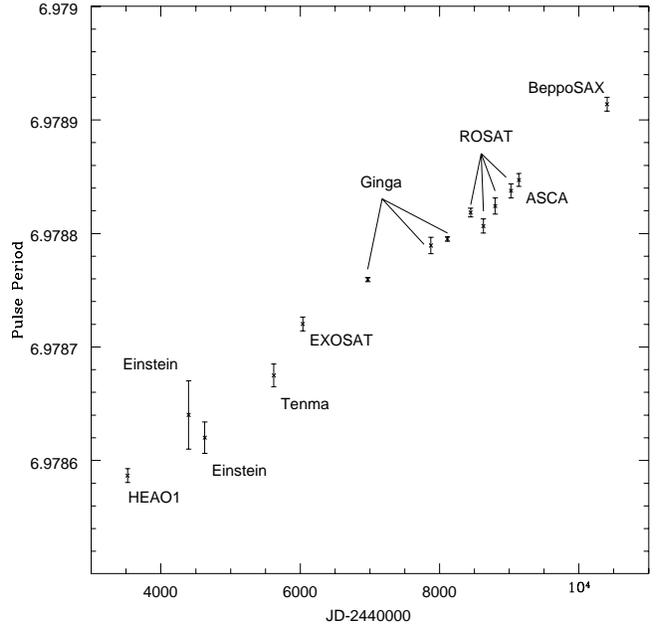,width=9.0cm}}
\caption[]{Pulse period history of \src. See Baykal \& Swank (1996) for
the measured values, except for the \sax\ result reported here}
\label{fig:spindown}
\end{figure}

\section {Discussion}
\label{subsec:discussion}

The region of sky containing the \src\ pulsar and the \snr\ SNR is
complex.  The \src\ spectrum is best described by the sum of a
power-law and a blackbody, confirming the results of Corbet et al.
(1995). These authors suggest that at least part of the power-law
component could originate from a synchrotron nebula which may be
visible around \src\ in \rosat\ images (Rho \& Petre 1997).  A similar
spectral decomposition has also been reported for two other
``anomalous pulsars'', 4U{\thinspace}0142+61 and
1E\thinspace1048.1$-$5937 (Whi\-te et al. 1996; Corbet \& Mihara 1997)
and has been interpreted as evidence for quasi-spherical accretion
onto isolated neutron stars after common envelope evolution and
spiral-in (Ghosh et al. 1997).  In this case the accretion flow has
two components.  A low-angular momentum component giving rise to the
blackbody emission from a large fraction of the neutron star surface,
and a high-angular momentum one forming an accretion disk responsible
for the power-law emission.

We caution that an alternative explanation for the observed
two-component spectrum cannot be excluded for \src. If a plausible
fraction of the emission arises from the part of the SNR within the
pulsar's extraction region, then an acceptable fit can be obtained
with a simple power-law model.  This implies that the derived
blackbody radius for \src\ should be regarded as an upper limit.  We
note that with the power-law fit, the pulsar's \nh\ of
$>$$2.1\times10^{22}$~\hcm\ is significantly greater than obtained for
\snr\ of $<$$7.6 \times 10^{21}$ \hcm.  This may indicate that the
pulsar lies some distance beyond the SNR and that the two objects are
unrelated, or it may indicate the presence of absorbing material local
to the pulsar. The \nh\ to the SNR is consistent with the galactic
column in the direction of \src\ of $7.4 \times 10^{21}$ \hcm\ (Dickey
\& Lockman 1990).
 
The derived pulse period is consistent with the extrapolation of the
long-term spin-down measured by Corbet et al. (1995) and Baykal \&
Swank (1996) as can be seen in Fig.~\ref{fig:spindown}.  No large
change in spin-down rate, as observed for example from
4U\thinspace1626$-$67 (Chakrabarty et al. 1997), is evident.  This
implies that the accretion torque has remained approximately constant
over at least the last 19~years.

In agreement with the results of Rho \& Petre (1997), the LECS
spectra of the SNR shell and of the jet-like X-ray lobe are 
indistinguishable, supporting a common origin for these two
regions.  We therefore refer to the summed spectrum as ``the SNR''.  
For an assumed distance
of 4~kpc, the 15\arcmin\ radius of the SNR
shell corresponds to 17.5\,pc, or an emitting volume of $f \times 6.6
\times 10^{59}\,$cm$^{3}$, where $f$ is the filling factor of the
emitting plasma within a sphere. The mass of
the SNR is then $60 \times f^{0.5}$~\Msun.
The canonical filling factor for a strong shock is 0.25, although the
absence of half of the remnant reduces this by a factor of 2. The
clumpiness of the X-ray emission (with most of the emission coming
from a few bright spots) implies an even smaller filling factor, and we
adopt a value of 0.1. The total emitting remnant mass is
therefore $\sim$15--20~\Msun. The remnant age is estimated 
to be 13,000~yr using the hydrodynamical model of Wang et al. (1992).
However this estimate uses a remnant plasma
temperature of 0.4\,keV, significantly
lower than determined here. The higher
temperature implies a faster expansion speed
(900\,km\,s$^{-1}$ rather than 590\,km\,s$^{-1}$),
and thus that the remnant is younger than previously estimated. The
Wang et al. (1992) estimate is based on detailed numerical simulations, and
it is not straightforward to determine the age corresponding 
to the updated temperature. Similarly,
Hughes et al. (1981) determine an age
of 17,000~yr assuming an X-ray temperature of
0.17\,keV. Given their analytical approach, is possible to 
scale their estimated age to the current temperature determination to
give a value of 3000~yr.

The low value derived for the emitting mass is consistent with a young
remnant, which has not yet swept up large quantities of
circum\-stellar material. The abundances derived from the X-ray
spectrum are slightly higher than cosmic values, indicative of
circumstellar material which has been mildly enriched either by the
stellar wind of the progenitor in its late evolutionary stages, or by
some moderate mixing-in of ejecta material.  The value of the
ionization parameter ($1.2 \times 10^3$\,cm$^{-3}$\,yr) is indicative
of strong non-equilibrium conditions, with an implied ionization age
of 3000\,yr, in good agreement with the scaled numerical simulations
of Hughes et al.  (1981). The lack of strongly enriched material
implies that the ejecta is not being directly observed, and is also
consistent with the lack of a strong wind of enriched gas originating
from the progenitor in its late stages. This is suggestive of a
low-mass progenitor and a Type~Ib supernova.

The LECS fit results to the whole of the SNR are similar to those
of Rho \& Petre (1997), who analyzed
BBXRT and \rosat\ spectra of a small part of the remnant,
just south of the pulsar. Their fit with a NEI model in
non-equipartition also shows strong non-equilibrium conditions, with
a comparable (considered the differences in the instrumental responses and 
plasma emission models) age of 6700~yr, 
near-cosmic or slightly
enriched abundances and an emitting mass of $85$~\Msun.

\begin{acknowledgements}
  The \sax\ satellite is a joint Italian--Dutch programme.
  T. Oosterbroek acknowledges an ESA Fellowship and S. Pightling a PPARC 
  studentship. Part of the data 
  reduction was carried out on the Southampton University Starlink node 
  which was funded by PPARC. We thank the referee, W. Brinkman, for
  helpful comments.
\end{acknowledgements}

\end{document}